# Room-temperature field-tunable radiofrequency rectification in epitaxial $SrIrO_3$ films

L. Zhou[1,†], Z. Z. Du[2,3,4,†], J. H. Wang[1], P. B. Chen[1], B. C. Ye[1], T. Feng[5], J. H. Yang[1], Z. H. Xiao[1], M. Yang[1], J.X. Li[1], W. Q. Zhang[5,6], H. Z. Lu[1,2,3,4,*] and H. T. He[1,6,7,*]

[1]*Department of Physics, Southern University of Science and Technology, Shenzhen 518055, China*
[2]*Shenzhen Institute for Quantum Science and Engineering, Southern University of Science and Technology, Shenzhen 518055, China*
[3]*Shenzhen Key Laboratory of Quantum Science and Engineering, Shenzhen 518055, China*
[4]*International Quantum Academy, Shenzhen 518048, China*
[5]*Department of Materials Science and Engineering, Southern University of Science and Technology, Shenzhen 518055, China*
[6]*Shenzhen Key Laboratory for Advanced Quantum Functional Materials and Devices, Southern University of Science and Technology, Shenzhen 518055, China*
[7]*Guangdong Provincial Key Laboratory of Advanced Thermoelectric Materials and Device Physics, Southern University of Science and Technology, Shenzhen, 518055, China*
[†]*These authors contributed equally: L. Zhou, Z.Z. Du*

[*]Corresponding author email address: luhz@sustech.edu.cn; heht@sustech.edu.cn

**Abstract**

Although significant advancements have been made in wireless technologies and portable devices, it remains a challenge for high-frequency and nanowatt-level radiofrequency rectification. In this work, we report a pronounced radiofrequency rectification up to 37 GHz in nominally centrosymmetric $SrIrO_3$ epitaxial films, with the minimum detectable power as low as ~300 nanowatts. Strikingly, the $SrIrO_3$ rectifier is highly field-tunable and exhibits a strong in-plane field anisotropy, thus showing a unique advantage in broad-band radiofrequency rectification. The rectification effect can persist up to at least 360 K and shows a sensitive temperature dependence including a sign inversion. By a systematic study of the nonlinear transport properties of $SrIrO_3$, it's further revealed that the radiofrequency rectification originates from the nonlinear Hall effect with the dominant contribution from field-induced Berry curvature dipole. Our work demonstrates the superior performance of the field-tunable $SrIrO_3$ rectifiers, unleashing the great application potential of centrosymmetric materials in harvesting and detecting ambient electromagnetic energy.

**Introduction**

Wireless radiofrequency(RF) rectification boosts flexibility and efficiency of energy transmission, greatly enhancing the integration of distributed electronic technology into industries and daily lives. With the rapid development of micro and nanoelectronics, high-frequency rectifiers are in high demand, particularly those operating in the microwatt and nanowatt range. Nevertheless, due to the limited electron transition time, the diode-based rectifiers exhibit poor RF performance at high frequencies (> GHz). At present, it is still challenging to achieve RF rectification in the frequency above 100 GHz. In view of this, the recent discovery of high-frequency rectification based on the nonlinear Hall effect (NLHE) has been attracting extensive attention [1, 2]. The NLHE, which generates a second-order Hall response to an alternating current in the absence of an external magnetic field, features two components, *i.e.*, a zero-frequency rectification component and a double-frequency component [3-7]. This rectification based on the NLHE emerges as a promising solution for high-frequency rectifiers, overcoming the inherent frequency limitations of semiconductor diodes [1,6]. However, earlier works about the NLHE appear predominantly at low temperatures and in non-centrosymmetric materials, such as bilayer or few-layer $WTe_2$ [8, 9], $MoTe_2$ [10], graphene [11, 12], $WSe_2$ [13, 14], $Bi_2Se_3$ [15] and $ZrTe_5$ [16], which greatly limits its application scenarios. Recently a few studies have demonstrated the room-temperature RF rectification in non-centrosymmetric $TaIrTe_4$ [17], BiTeBr [18], Te [19] and $BaMnSb_2$ [20]. But the first two are exfoliated mechanically from crystals and the last two are nanoflakes or bulk crystals, which would hinder integration into large-scale applications. Encouragingly, applying an external DC electric field could induce a NLHE in centrosymmetric materials like bulk $WTe_2$ [21, 22], but an efficient room-temperature RF rectification is still absent in the literature.

In this work, we demonstrate successfully a field-induced RF rectification up to at least 37 GHz at room temperature in centrosymmetric $SrIrO_3$ epitaxial films. In particular, the rectifier exhibits high magnitude and sign tunability to the DC electric field, with the resultant rectified voltage about one order of magnitude larger than previously reported values in diode-free rectifiers [18]. By in-plane field rotation, we further reveal a strong field anisotropy of this RF rectification. It's found that the rectified response is greatly enhanced once the field direction is parallel with the mirror lines of $SrIrO_3$. Due to a temperature-driven shift in the Fermi energy, a sensitive temperature dependence is also observed for the rectification, featuring a sign inversion at about 220 K. A comparative study of this RF rectification and the NLHE leads us to believe that the rectification arises from the NLHE with dominant intrinsic contribution from field-induced Berry curvature dipole. Our work not only reveals the efficient field-tunable rectification of ambient RF signals in $SrIrO_3$ epitaxial films, with a unique advantage for future large-scale broad-band applications, but also paves the way for other centrosymmetric materials to be utilized in RF rectifiers or detectors.

**Results and Discussions**

The 5$d$ transition metal oxide SrIrO$_3$, belonging to the centrosymmetric *Pbnm* space group, has been shown to exhibit a nonlinear planar Hall effect within an in-plane magnetic field [23-27], but no rectification or NLHE has been addressed. It is worth pointing out that there exists a mirror symmetry plane and a glide mirror symmetry plane in the epitaxial SrIrO$_3$ film, which are perpendicular to the $(1\bar{1}0)$ and (001) crystal directions, respectively [28-30], as indicated in Fig. 1(a). Figure 1(b) shows schematically the device structure to investigate the RF rectification in SrIrO$_3$. The device is exposed to a RF radiation (*e.g.*, at 1 GHz) with the DC electric field $E^{DC}$ applied along the direction $(1\bar{1}0)$ and the rectification voltage $V_H^{Rect}$ measured in the same direction. Note that in obtaining $V_H^{Rect}$, the background DC voltage due to $E^{DC}$ has been subtracted.

As shown in Fig. 1(c), in the absence of an external DC electric field $E^{DC}$, the RF rectification effect at 300 K is negligible, *i.e.* $V_H^{Rect} \sim 0$, regardless of the input RF power ($P$). However, upon applying $E^{DC}$, a pronounced RF rectification response is observed, with the rectification voltage $V_H^{Rect}$ linearly dependent on the input RF power. The rectification voltage $V_H^{Rect}$ can be effectively modulated by the applied $E^{DC}$, *i.e.*, the magnitude of $V_H^{Rect}$ increases as stronger electric fields are applied and the sign changes once the field direction is reversed. Additionally, as shown in Fig. 1(d), the RF rectification effect can occur up to 37 GHz, which not only surpasses the traditional silicon diode cut-off frequency [31], but also shows potential applications in Wi-Fi (2.4 GHz, 5.9 GHz), UAV communications (1-40 GHz), radar systems (1-12 GHz), and satellite communications (12-40 GHz). Remarkably, the voltage responsivity $R_V$ defined as the slope of the $V_H^{Rect}(P)$ curve can reach up to at least 0.22 V/W at 5.9 GHz, one order of magnitude larger than the largest value reported so far in rectifiers based on the intrinsic NLHE (~0.02 V/W @ 5.9 GHz in BiTeBr [18], table S1 & Fig. S1(a) [32]). Moreover, the RF rectification of SrIrO$_3$ can be observed with the RF power down to -15 dBm (~300 nW), as shown in Fig. S1(b) [32]. This power threshold falls within the ambient RF power range between -20 and -10 dBm [31], indicating the rectifier's capability to rectify the ambient RF signals. Most importantly, due to the field-tunability shown in Fig. 1(c), or the linear field dependence of $R_V$ shown in Fig. S2 [32], the SrIrO$_3$ rectifier can even be electrically tuned to have the same $R_V$ when subjected to RF excitations with different frequencies. As shown in Fig. 1(e), by setting appropriate $E^{DC}$, the obtained $V_H^{Rect}(P)$ curves coincide with each other, yielding the same value of $R_V$. Note that for a comparison between the results shown in Fig. 1(d) and 1(e), the electric field condition is kept the same for 5.9 GHz. This field-tunability thus gives the SrIrO$_3$ rectifier a unique advantage over other rectifiers in broad-band RF rectifications [38]. All the above results collectively demonstrate the superior performance of RF rectification in SrIrO$_3$, suggesting the great application potential in harvesting and detecting high-frequency ambient electromagnetic energy.

We have further investigated the field anisotropy of the RF rectification effect. To do so, we have prepared a disk-shaped multi-terminal device as illustrated in Fig. 2(a). An alternating electric field $E^\omega$ at 1 GHz is applied in the (001) direction. We measure the rectification voltage $V_H^{Rect}$ in the ($1\bar{1}0$) direction while varying the direction of $E^{DC}$. Here, the angle between $E^{DC}$ and (001) is denoted as $\theta$ and the magnitude of $E^{DC}$ is fixed at 20.3 kV/m. Figure 2(b) shows that the field-induced RF rectification is negligible when the DC field is along the (001) direction ($\theta=0°$ & $180°$), the same as the applied alternating electrical field $E^\omega$. As the DC field is titled away from the (001) direction, the $V_H^{Rect}$ is enhanced. The rectification effect is maximized when the field is perpendicular to the (001) direction, i.e., $\theta=90°$ & $270°$. We also swap the directions of $E^\omega$ and $V_H^{Rect}$, i.e., $E^\omega \parallel (1\bar{1}0)$ & $V_H^{Rect} \parallel (001)$, and repeat the above measurement. As shown in Fig. 2(c), the $V_H^{Rect}$ is minimized at $E^{DC} \parallel (1\bar{1}0)$ ($\theta=90°$ & $270°$), but maximized at $E^{DC} \parallel (001)$ ($\theta=0°$ & $180°$), opposite to the results in Fig. 2(b). To see more clearly the field direction dependence of the RF rectification, we plot the values of $R_V$ as a function of the field angle $\theta$ for both $E^\omega$ directions ($E^\omega \parallel (001)$ or $E^\omega \parallel (1\bar{1}0)$) in Fig. 2(d). A strong field anisotropy of the rectification effect is clearly observed, revealing the essential role of crystal symmetry in this phenomenon.

Besides anisotropy of this rectification, we have also studied the influence of temperature on the field-induced RF rectification effect in SrIrO$_3$. Figure 3(a) shows the $V_H^{Rect}(P)$ curves at different temperatures, with $E^\omega \parallel (001)$ & $E^{DC} \parallel (1\bar{1}0)$ to maximize the rectification effect. The rectification voltage $V_H^{Rect}$ remains linearly dependent on the input power $P$ at any measured temperature. But the slope of the $V_H^{Rect}(P)$ curve varies with changing temperature. Especially there exists a temperature-induced sign change in the slope. Figure 3(b) shows the temperature dependence of the rectification voltage responsivity $R_V$, obtained by linear fitting of the curves in Fig. 3(a). The $R_V$ exhibits a monotonically increasing behavior as the temperature increases. Around 220 K, a sign inversion from negative to positive is observed. Note that similar phenomena have also been observed in other SrIrO$_3$ rectifiers (Fig. S3 [32]). The RF rectification can persist up to at least 360 K, which has not yet been reported in previous works about rectifiers based on the NLHE [17-20]. These results clearly show the sensitive temperature dependence of the RF rectification in SrIrO$_3$ and its potential in high temperature applications.

The linear dependence of rectification voltage on the input power, as well as its strong anisotropy, align well with the NLHE [1,21]. Though the NLHE generally occurs in non-centrosymmetric systems, previous research has shown that applying an external DC electric field can also induce a NLHE in centrosymmetric materials due to nonzero Berry connection polarizability, such as bulk WTe$_2$ [21]. In order to investigate whether the observed RF rectification originates from the NLHE, we have measured the two components of NLHE simultaneously, i.e., the rectified and second-harmonic Hall voltages $V_H^{Rect}$ & $V_H^{2\omega}$, at a frequency of 17.7 Hz. The results are shown in Fig. 4(a) where the alternating current $I^\omega$ was applied in the (001) direction and the DC electric field $E^{DC}=20.3$ kV/m in the ($1\bar{1}0$) direction. Both the measured $V_H^{Rect}$ and $V_H^{2\omega}$ show a quadratic dependence on $I^\omega$, as indicated by the red fitting curves with $V_H = \alpha(I^\omega)^2$. It is also found that $V_H^{Rect} \sim 1.44 V_H^{2\omega}$, in agreement with a previous experimental study of the NLHE in BaMnSb$_2$ [20]. Considering that the $V_H^{2\omega}$ measured by a lock-in amplifier is its root-mean-square value, the above finding indicates that the amplitude of $V_H^{2\omega}$ is almost equal to $V_H^{Rect}$, consistent with the theoretical expectations [3]. Like the RF rectification, the DC electric field is also crucial to the observation of second harmonic Hall voltage. $V_H^{2\omega} \sim 0$ if no DC electric field $E^{DC}$ is applied (Fig. S4(a) [32]). This is expected as there is no inversion symmetry breaking in our SrIrO$_3$ films. But in the presence of a DC electric field, we can detect an obvious $V_H^{2\omega}$. All the measured $V_H^{2\omega}$ shows a quadratic dependence on $I^\omega$ (Fig. S4(a) [32]). We have also studied the field anisotropy of the $V_H^{2\omega}$ at low frequency (Fig. S4 [32]), which is consistent with the results of the RF rectification shown in Fig.2. Therefore, the simultaneous observation of the quadratic $V_H^{Rect}$ and $V_H^{2\omega}$ with equal magnitudes and the consistent field anisotropy of the nonlinear response strongly indicate that the RF rectification is intimately linked to the NLHE in SrIrO$_3$.

Generally, the NLHE obeys the scaling law $\chi_{yxx}^{(2)} = (L^3/W^2 dR_L^3)[V_H^{2\omega}/(I^\omega)^2] = \eta_1 \tau + \eta_2 \tau^2 + \eta_3 \tau^3$, where $L$ is the effective length of the longitudinal signal, $W$ is the effective length of the traverse signal, $d$ is the thickness of the sample, $\tau$ is the scattering time, and the parameter $\eta_i$ characterizes the different contributions to the second-order nonlinear Hall conductivity $\chi_{yxx}^{(2)}$ [22, 39, 40]. The $\eta_1$ term scaling as $\tau$ might originate from the Berry curvature dipole or extrinsic scattering [4, 9, 15, 39, 41]. The extrinsic scattering, including the side jump and skew scattering, can also give rise to the $\eta_2$ term which scales as $\tau^2$ [39]. The last term is usually ascribed to the Drude transport or skew scattering [17, 39]. Therefore, based on the nonlinear Hall data and scattering time $\tau$ obtained at different temperatures, one can perform the scaling law analysis to evaluate the different contributions to the NLHE. Such analysis has been done in previous studies of the NLHE in T$_d$-MoTe$_2$ [10], bilayer graphene [11], Ce$_3$Bi$_4$Pd$_3$ [42], BaMnSb$_2$ [20], Mn$_2$BiTe$_4$ [40] *etc.*, as well as the field-induced NLHE in WTe$_2$ [20,21].

We have also analyzed the dependence of the second-order conductivity $\chi_{yxx}^{(2)}$ on the scattering time $\tau$, as shown in Fig. 4(b). The $\chi_{yxx}^{(2)}$ is obtained using the formula $\chi_{yxx}^{(2)} = (L^3/W^2 dR_L^3)[V_H^{2\omega}/(I^\omega)^2]$ and $\tau$ is derived with the formula $\tau = m^*\sigma/ne^2$, where $m^*$ is the effective mass [37], $n$ is the carrier concentration, and $\sigma$ is the electrical conductivity. Based on the ordinary Hall measurement, the derived values of $\tau$ & $n$ at different temperatures are shown in Fig. S5 [32]. As seen in Fig. 4(b), the $\chi_{yxx}^{(2)}$ initially shows a linear dependence on $\tau$, as indicated by the black dashed line. Note that there also occurs a sign change in $\chi_{yxx}^{(2)}$. As the scattering time increases, $\chi_{yxx}^{(2)}$ gradually deviates from this linear trend, and shows an upturn at high $\tau$ values. The linear dependence of $\chi_{yxx}^{(2)}$ on $\tau$ indicates the dominant intrinsic contribution to the field-induced NLHE in our SrIrO$_3$ epitaxial films [21, 43, 44]. As for the sign inversion in $\chi_{yxx}^{(2)}$, we recall that similar phenomena have been observed in TaIrTe$_4$ and ascribed to the change in the intrinsic contribution due to a temperature-induced shift in the Fermi level [17]. Since the carrier density in Fig. S5 [32] shows an obvious decrease with decreasing temperatures, *i.e.*, the Fermi level shifts down as temperatures decrease, we thus believe that the same mechanism is also applicable to SrIrO$_3$. It's worth pointing out that almost at the same temperature, the rectification voltage responsivity $R_V$ also exhibits a sign change, as shown in Fig. 3(b). Therefore, the observation of sign changes in $\chi_{yxx}^{(2)}$ and $R_V$ once again points to the close connection between the NLHE and the RF rectification. To understand the deviation of $\chi_{yxx}^{(2)}$ from the linear behavior at high $\tau$ values, we notice that the minimum of $\chi_{yxx}^{(2)}$ occurs at about 110 K by comparing Fig. 4(b) and Fig. S5 [32]. As revealed in previous studies [45], the SrTiO$_3$ substrate endures a cubic-to-tetragonal phase transition around 110 K. This coincidence thus suggests that the structural phase transition in the SrTiO$_3$ substrate could influence the NLHE of the SrIrO$_3$ films epitaxially grown on the substrate [16]. Compared with $\chi_{yxx}^{(2)}$, other transport parameters, such as $n$ & $\tau$ in Fig. S5 [32], show no obvious change around 110 K. Therefore, the NLHE is far more sensitive to the crystal symmetry, which might be important to the crystal symmetry characterization in various materials [6].

Note that the observed NLHE is not due to the diode rectification effect as Ohmic contacts to the SrIrO$_3$ film with linear current-voltage characteristics are obtained in our experiment [32]. The anisotropy of the NLHE also suggests the minor role of thermal or thermoelectric effects in generating the phenomena [32]. But for a full understanding of the above-discussed nonlinear transport phenomena in SrIrO$_3$, a first-principle calculation of Berry curvatures and its comparison with the experimental results are highly desired in the near future.

**Conclusions**

Field-tunable RF rectification up to 37 GHz has been successfully realized in centrosymmetric epitaxial SrIrO$_3$ thin films at temperatures as high as 360 K and ascribed to the field-induced NLHE. The rectifier exhibits superior rectification performance and shows a unique advantage in broad-band applications due to the field tunability. Therefore, our work would shed new light on the utilization of centrosymmetric materials in harvesting and detecting ambient electromagnetic energy. Different from previous studies where room-temperature RF rectifications were reported in micrometer-sized flakes mechanically exfoliated from bulk crystals or fabricated by focused ion beam [17-20], the SrIrO$_3$ films studied in this work are epitaxially grown on centimeter-sized substrates, which is thus of particular importance in large-scale fabrication and integration. Besides, the perovskite oxide family has been known to host a broad spectrum of functional properties, such as the ferroelectricity, piezoelectricity, pyroelectricity, and magnetism [46]. The epitaxial integration of SrIrO$_3$ thin films with other perovskite oxides, together with the efficient field-tunable RF rectification observed in this work, might offer abundant possibilities for future multifunctional device designs.

**Figures**

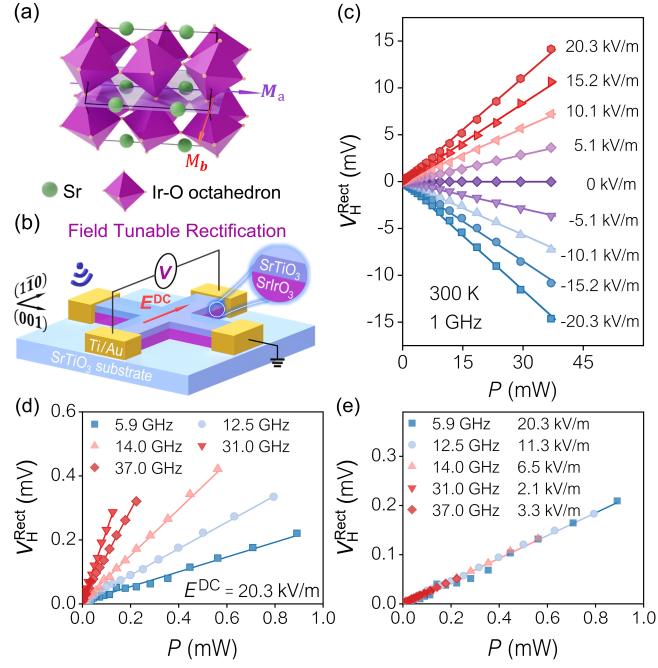

**FIG. 1.** (a) Schematic crystal structure of SrIrO$_3$, with the arrows indicating the two mirror lines $M_a$ & $M_b$. (b) Measurement setup for the RF rectification voltage $V_H^{Rect}$ induced by a DC electric field $E^{DC}$. (c) With the frequency fixed at 1 GHz, the $V_H^{Rect}(P)$ curves measured under different $E^{DC}$ conditions. (d) With $E^{DC}$ = 20.3 kV/m, the $V_H^{Rect}(P)$ curves measured under different RF conditions. (e) The $V_H^{Rect}(P)$ curves for different frequencies overlap each other by setting appropriate $E^{DC}$.

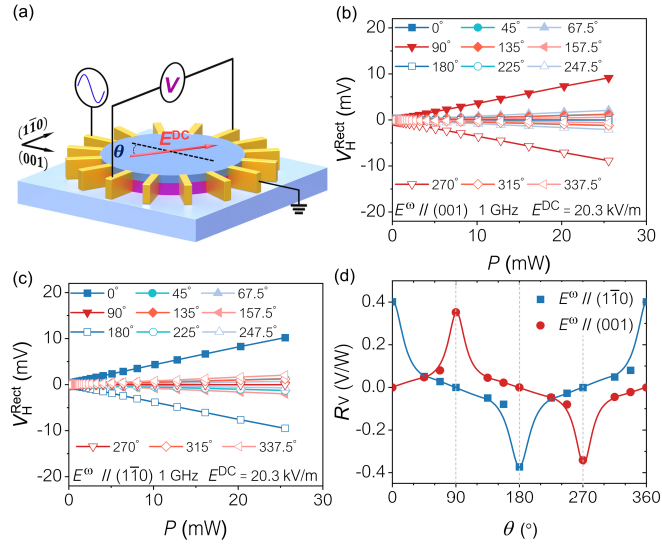

**FIG. 2.** (a) Device layout and measurement configuration for studying the rectification anisotropy. The angle between the DC electric field $E^{DC}$ and the (001) crystal direction is denoted as $\theta$. The magnitude of $E^{DC}$ and radiation frequency are fixed at 20.3 kV/m and 1 GHz, respectively. (b) With

$E^\omega \parallel (001)$ & $V_H^{Rect} \parallel (1\bar{1}0)$, the measured $V_H^{Rect}(P)$ curves under different $\theta$ conditions. (c) With $E^\omega \parallel (1\bar{1}0)$ & $V_H^{Rect} \parallel (001)$, the measured $V_H^{Rect}(P)$ curves under different $\theta$ conditions. (d) Derived rectification voltage responsivity $R_V$ for different DC electric field directions with $E^\omega \parallel (1\bar{1}0)$ & $(001)$, respectively. The solid lines in the Fig. 2(d) are guide for the eye.

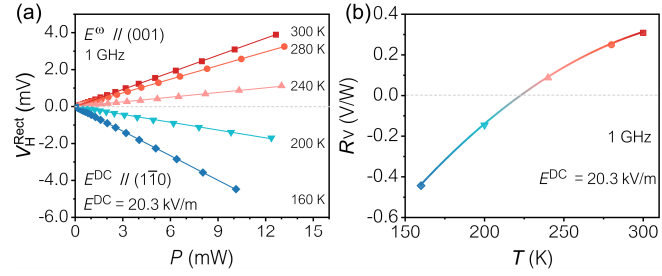

**FIG. 3.** (a) The $V_H^{Rect}(P)$ curves measured at different temperatures. To maximize the rectification effect, the $E^\omega$ at 1 GHz is applied in the (001) direction, while with $E^{DC}$=20.3 kV/m in the $(1\bar{1}0)$ direction. (b) The derived rectification voltage responsivity $R_V$ as a function of temperature.

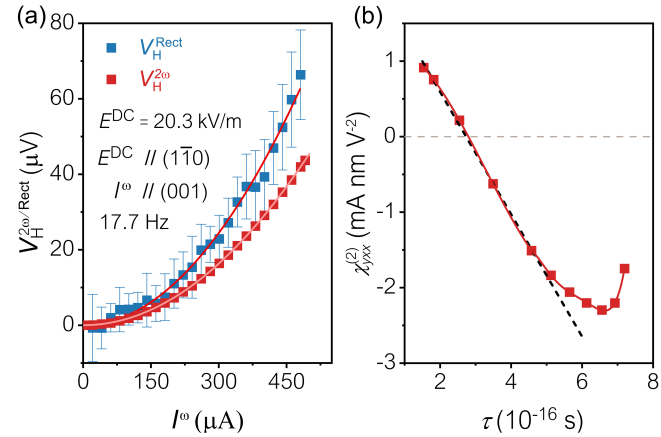

**FIG. 4.** (a) The AC current $I^\omega$ dependence of the field-induced rectification and second-harmonic Hall voltages $V_H^{Rect}$ & $V_H^{2\omega}$ at $T$= 300 K. The AC current with the frequency of 17.7 Hz and the DC electric field $E^{DC}$= 20.3 kV/m are applied in the (001) and $(1\bar{1}0)$ directions, respectively. Both curves can be well fitted with $V_H = \alpha(I^\omega)^2$. (b) The scattering time $\tau$ dependence of the nonlinear second-order conductivity $\chi_{yxx}^{(2)}$.

## Acknowledgment

L. Z. and Z. Z. D. contributed equally to this work. This work was supported by the National Key Research and Development Program of China (2022YFA1403700), the Natural Science Foundation